\documentclass[aps,twocolumn,showpacs]{revtex4-1}
\usepackage{float}
\usepackage{makeidx}
\usepackage{amsmath,amssymb,graphics,epsfig,color,times,bbm}

\begin{document}

\title{Cascaded Entanglement Enhancement}
\author{Zhihui Yan}
\author{Xiaojun Jia}
 \email{jiaxj@sxu.edu.cn}
\author{Xiaolong Su}
\author{Zhiyuan Duan}
\author{Changde Xie}
\author{Kunchi Peng}

\affiliation{State Key Laboratory of Quantum Optics and Quantum Optics
Devices, Institute of Opto-Electronics, Shanxi University, Taiyuan, 030006,
P. R. China}

\begin{abstract}
We present a cascaded system consisting of three non-degenerate optical
parametric amplifiers (NOPAs) for the generation and the enhancement of
quantum entanglement of continuous variables. The entanglement of optical
fields produced by the first NOPA is successively enhanced by the second and
the third NOPAs from $-5.3$ $dB$ to $-8.1$ $dB$ below the quantum noise
limit. The dependence of the enhanced entanglement on the physical
parameters of the NOPAs and the reachable entanglement limitation for a
given cascaded NOPA system are calculated. The calculation results are in
good agreement with the experimental measurements.
\end{abstract}


\maketitle

Nonlocal quantum entanglement is the key resource to realize quantum
information processing (QIP) \cite{Galindo,Braunstein,Reid}. The entangled
states of single photons (qubits) and optical modes (qumodes) have been
applied in QIP with discrete and continuous variable (DV and CV) regimes,
respectively \cite{Bouwmeester,Furusawa}. The quadrature squeezed states of
light are the essential resource states in CV QIP since squeezing is the
necessary base to establish quantum entanglement among optical fields \cite
{Wu}. A scheme of generating CV optical entangled states is to interfere two
single-mode squeezed states of light with an identical frequency and a
constant phase-difference on a $50/50$ beamsplitter \cite{Furusawa,Lance}. The two
single-mode squeezed states are often produced by a pair of degenerate
optical parametric amplifiers (DOPAs) with identical type-I nonlinear
crystal pumped by a laser to ensure high interference efficiency. Through
carefully technical improvement on suppressing the phase fluctuation of
optical field and reducing the intra-cavity losses of DOPA, the squeezing
level of the single-mode squeezed states is continually renewed in recent
years \cite{Eberle,Mehmet,Takeno}. So far, the squeezing over $-12$ $dB$ below the
quantum noise limitation (QNL) has been achieved by a group in Hannover \cite
{Eberle,Mehmet}. Coupling a single-mode squeezed state of $-9.9$ $dB$ and a
vacuum field on a $50/50$ beam splitter, the Einstein-Podolsky-Rosen (EPR)
entangled state of light (also named two-mode squeezed state) with the
quantum correlation of amplitude and phase quadratures of $-3$ $dB$ below the QNL was
obtained in 2011 \cite{Eber}. The four-mode CV entanglement of $-6$ $dB$
below the QNL was achieved by combing four initial single-mode squeezed
states generated by four DOPAs \cite{Yukawa}. The highest CV entanglement
produced by coupling single-mode squeezed states is $-6$ $dB$ below the QNL up to
now \cite{Yukawa}.

Another important device to generate CV EPR\ entangled state of optical
field is the non-degenerate optical parametric amplifier (NOPA) consisting
an optical cavity and a type-II nonlinear crystal. Through the intra-cavity
frequency-down conversion process in a NOPA, a pair of non-degenerate optical
modes with amplitude and phase quadrature correlations is directly produced,
which is an EPR entangled state \cite{Ou}. Twenty years ago, Kimble's group
experimentally generated a pair of CV entangled optical beams with a NOPA
and demonstrated the EPR paradox firstly in CV regime \cite{Ou}. Then, NOPAs
operating at different version (above or lower the threshold, amplification
or de-amplification) are used as the sources of generating optical CV
entangled states and the produced EPR beams are applied in a variety of CV\
QIP experiments \cite{Li,Jia,Laurat,Villar,Jing,Keller}. However, in a long
period the EPR entanglement level was kept around $-4$ $dB$ or lower \cite
{Ou,Li,Jia,Laurat,Villar,Jing,Keller}. Until 2010, after a series of strictly
technical improvement on the NOPA system, the EPR entanglement degree was
raised to $-6$ $dB$ \cite{Wang}, which is the best reported result on NOPA
system.

In order to obtain high levels of squeezing and entanglement using a single DOPA or NOPA, the device has to be optimized to achieve low phase fluctuations
and low optical losses. Especially, a nonlinear optical crystal with a high
second-order nonlinear coefficient and low loss is desired. However, the
quality of nonlinear crystals and optical elements is not ideal
usually. In the quantum manipulation experiments of CV entangled states, we
found that the EPR entanglement of optical field produced by an NOPA can be
enhanced by another NOPA under appropriate operation condition \cite
{Chen,Shang}. The amount of the enhanced entanglement is limited by the
physical parameters of the used NOPA. For further raising entanglement under
generally technical condition, we design the cascaded NOPA system
involving three NOPAs and experimentally realized the cascaded amplification
of CV entanglement. The initial EPR entanglement of $-5.3$ $dB$ produced by
the first NOPA (NOPA1) is enhanced to $-7.2$ $dB$ by the second one (NOPA2)
and successively to $-8.1$ $dB$ by the third one (NOPA3), which is the
highest EPR entanglement of optical modes obtained by experiments so far to
the best of our knowledges. We numerically calculate the correlation
variances of the enhanced EPR entangled state based on the physical
parameters of the experimental system. The calculated results are in good
agreement with the experimental measurements.

\begin{figure}[tbp]
\begin{center}
\includegraphics[width=8.6cm]{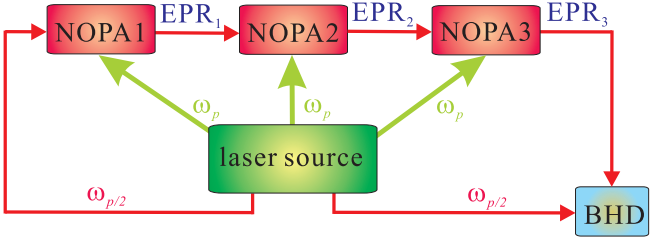}
\end{center}
\caption{(Color online) The principle schematic of the cascaded entanglement
enhancement system. }
\end{figure}

Fig.1 shows the principle schematic of the cascaded entanglement enhancement
system. The laser source is an intra-cavity frequency-doubled CW laser, the
output second-harmonic wave ($\omega _p$) from which is used for the pump
fields of the three NOPAs and the output fundamental wave ($\omega _{p/2}$)
serves as the injected signal of NOPA1 and the local oscillator (LO) of the
balanced homodyne detector (BHD) for the entanglement measurement. The EPR
entangled light generated by NOPA1 (EPR1) is injected into NOPA2 as the
injected signal for the first-stage enhancement of the entanglement and the
the amplified EPR optical field (EPR2) is injected into NOPA3 for the
second-stage enhancement. The final entangled light (EPR3) is detected by
the BHD. To achieve the optimal entanglement enhancement, the three NOPAs
should be operated at an identical state \cite{Chen,Shang}. In the presented
experiment, the three NOPAs are operated below the oscillation threshold of
NOPA and at the de-amplification state, i.e. the pump field and the injected
signal are out of phase (with the phase difference of $(2n+1)\pi $, $n$%
-integer). We also require that the signal and idler modes obtained through
the frequency-down-conversion of the pump field inside the NOPA have an
identical frequency ($\omega _{p/2}$) and the orthogonal polarization, which
is easily satisfied in experiments \cite{Ou,Li,Jia,Laurat}. In this case the
produced entangled states have the correlated amplitude-sum and
phase-difference as well as the anti-correlated amplitude-difference and
phase-sum \cite{Li,Jia,Laurat}.

In the following we calculate the correlation variances between the
amplitude and phase quadratures of the EPR entangled state enhanced by a
NOPA firstly. We describe the quantum state of light with the
electromagnetic field annihilation operators $\hat{a}=(\hat{X}+i\hat{Y})/2$,
where $\hat{X}$ and $\hat{Y}$ are the operators of the amplitude ($\hat{X}$)
and the phase ($\hat{Y}$) quadratures, respectively. $\hat{X}$ and $\hat{Y}$
satisfy the canonical commutation relation $[\hat{X},\hat{Y}]=2i$. For
generating high EPR entanglement in experiments, the operation conditions of
the subharmonic signal ($\hat{a}_1$) and idler ($\hat{a}_2$) modes in NOPA
should be balanced, i. e. they have the same transmissivity efficiency ($%
\gamma _1$) on the input-output coupler of the optical cavity (Supposing that a cavity mirror serves as the input and the output coupler simultaneously. See Fig. 3) and the
identical intra-cavity loss ($\gamma _2$). The correlation variances and the
anti-correlation variances of the injected EPR optical modes, which is
produced by the former NOPA, are expressed by $\langle \delta ^2(\hat{X}%
_{a_{_1}}^{in}+\hat{X}_{a_{_2}}^{in})\rangle =\langle \delta ^2(\hat{Y}%
_{a_{_1}}^{in}-\hat{Y}_{a_{_2}}^{in})\rangle =2e^{-2r}$ and $\langle \delta
^2(\hat{X}_{a_{_1}}^{in}-\hat{X}_{a_{_2}}^{in})\rangle =\langle \delta ^2(%
\hat{Y}_{a_{_1}}^{in}+\hat{Y}_{a_{_2}}^{in})\rangle =2e^{2r+2r^{\prime }}$,
where $r$ and $r^{\prime }$ are the correlation parameter and the extra
noise factor on the anti-correlation components, respectively; $\hat{X}%
_{a_{_{1(2)}}}^{in}$ and $\hat{Y}_{a_{_{1(2)}}}^{in}$ stand for the
amplitude and the phase quadratures of the injected mode $\hat{a}_{1(2)}^{in}$,
respectively \cite{Chen,Zhang}. Solving the quantum Langevin motion equations
and using the input-output relation of the NOPA, the correlation variances
of the output field are obtained:

\begin{eqnarray}
&&\langle \delta ^2(\hat{X}_{a_{_1}}^{out}+\hat{X}_{a_{_2}}^{out})\rangle
=\langle \delta ^2(\hat{Y}_{a_{_1}}^{out}-\hat{Y}_{a_{_2}}^{out})\rangle
\nonumber \\
&=&(\zeta (2\frac{(-\kappa +\gamma _1-\gamma _2)^2+(\omega \tau )^2}{(\kappa
+\gamma _1+\gamma _2)^2+(\omega \tau )^2}e^{-2r}  \nonumber \\
&&+2\frac{(2\sqrt{\gamma _1\gamma _2})^2}{(\kappa +\gamma _1+\gamma
_2)^2+(\omega \tau )^2})+1-\zeta )\cos ^2\theta  \nonumber \\
&&+(\zeta (2\frac{(-\kappa +\gamma _1-\gamma _2)^2-(\omega \tau )^2}{(\kappa
+\gamma _1+\gamma _2)^2+(\omega \tau )^2}e^{2r+2r^{\prime }}  \nonumber \\
&&+2\frac{(2\sqrt{\gamma _1\gamma _2})^2}{(\kappa +\gamma _1+\gamma
_2)^2+(\omega \tau )^2})+1-\zeta )\sin ^2\theta ,
\end{eqnarray}
where $\hat{X}_{a_{_{1(2)}}}^{out}$ and $\hat{Y}_{a_{_{1(2)}}}^{out}$ are
the amplitude and the phase quadratures of the output mode $\hat{a}%
_{1(2)}^{out}$, respectively; $\kappa =\beta \chi $ is the nonlinear
coupling efficiency of the NOPA which is proportional to the pump parameter $%
\beta =(p_{pump}/p_{th})^{1/2}$ ($p_{pump}$ - the pump power, $p_{th}$ - the
threshold pump power of NOPA) and the second-order nonlinear coupling
coefficient $\chi $ of the nonlinear crystal used in the NOPA; $\tau $ is
the round-trip time of light in the optical cavity; $\omega =2\pi \Omega $
is the noise analysis frequency; $\zeta $ is the imperfect detection
efficiency; and $\theta $ is the relative phase fluctuation between the pump field and the injected signal resulting from imperfect phase-locking.

\begin{figure}[tbp]
\begin{center}
\includegraphics[width=8.6cm]{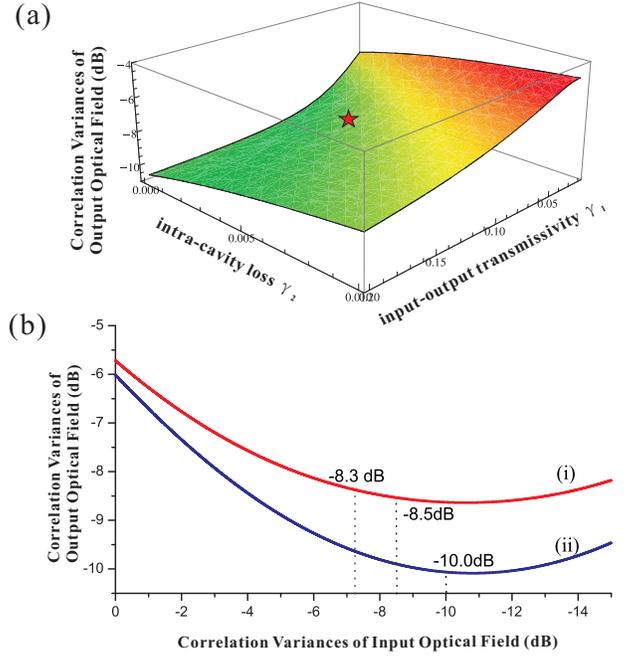}
\end{center}
\caption{(Color online) (a) The dependence of correlation variances of the
output entangled optical field on the input-output transmissivity $\gamma
_{1}$ and the intra-cavity loss $\gamma _{2} $ of NOPA. (b) The functions of the correlation
variances of the output field vs that of the input field for two NOPAs: (i) $\gamma _1=0.1$, $\gamma _2=0.004$; (ii) $\gamma _1=0.1$, $\gamma _2=0.001$. }
\end{figure}

Fig. 2 (a) is the calculated dependence of correlation variances of the
output EPR optical field from NOPA3 on the cavity parameter $\gamma _1$ and $%
\gamma _2$, where other parameters are taken according to the really
experimental system ($r=0.83$; $r^{\prime }=0.45$; which correspond to the entanglement degree of EPR2 with correlation variance of -7.2 dB below the QNL and the anti-correlation variance of 11.1 dB above the QNL, see the experimental results in the text; $\theta =0.0105;\Omega =2.0$ $MHz;\tau =2.0*10^{-9}$ $s;\varsigma =0.947.$). When $\gamma _1$ increases and $%
\gamma _2$\ decreases the correlation variance of the output field reduces,
i.e. the entanglement degree increases. It means that for a simple NOPA the
higher input-output transmissivity ($\gamma _1$) and the lower intra-cavity
loss ($\gamma _2$) can provide the stronger performance of the entanglement
enhancement. However, for a NOPA\ with given physical parameters, the
correlation variances of the output field depend on the correlation
variances of the injected signal field. Fig. 2 (b) shows the functions of
the correlation variances of the output field vs that of the input field,
where the trace (i) for $\gamma _1=0.1$, $\gamma _2=0.004$ and the trace
(ii) for $\gamma _1=0.1$, $\gamma _2=0.001$; other parameters are the same
as that of Fig. 2 (a). We can see from Fig. 2 (b), there is a turning point
in the trace (i) ($-8.5$ $dB$) and (ii) ($-10.0dB$),
respectively, where the correlation variance of the output field equals to that of the input field. And after the turning point the correlation variances of the output field will be larger than that of the input signal. It means the enhancement ability of the NOPA no longer exists. Comparing traces (i) and (ii), it is obvious that the NOPA with the
smaller intra-cavity loss (trace (ii)) has stronger ability of the
entanglement enhancement. For a given NOPA, the ability of the entanglement
enhancement decreases when the entanglement of the injected field increases.
After the turning point of the trace (i) and (ii), the entanglement of the
output field becomes worse than that of the input field, i.e. there is an
upper limitation of the entanglement of the input field for a given NOPA. When the entanglement of the input field surpasses the limitation value, the
entanglement of the output field starts to reduce. This is because the
noises of the anti-correlation components in the input field increase rapidly along with the
increasing of the correlation degree, and the noises on the anti-correlation components would be inevitably coupled to the correlation component and thus decrease the correlation degree of the output fields due to the existence of the phase fluctuation $\theta$ in the experimental system (See the second term of Eq. (1)). Reducing the phase fluctuation in the phase-locking system, the upper limitation of the input field for the entanglement enhancement will be raised.

\begin{figure}[tbp]
\begin{center}
\includegraphics[width=8.6cm]{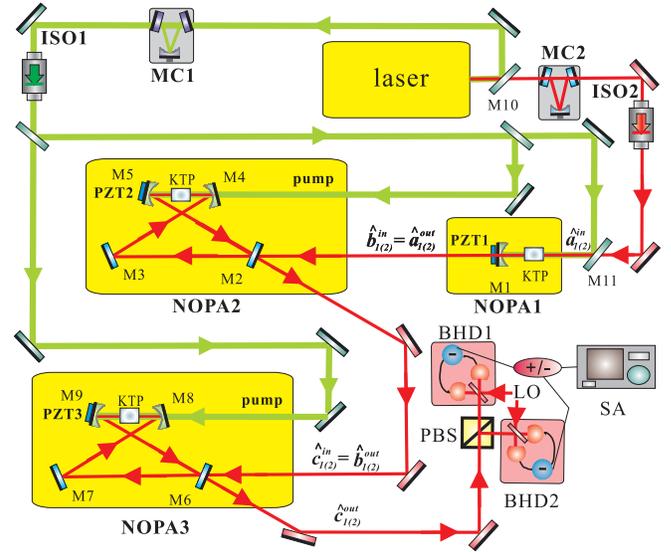}
\end{center}
\caption{(Color online) The experimental setup of cascade sensitive-phase
entanglement enhancement. laser, Nd:YAP/LBO laser source; PBS, polarizing
beam splitter; MC1(2), mode cleaner; ISO1(2), optical isolator; M0--M11,
different mirrors (see text for details); PZT, piezoelectric transducer;
+/-, positive/negative power combiner; SA, spectrum analyzer.}
\end{figure}

The experimental setup is shown in Fig.3. A continuous-wave intra-cavity
frequency-doubled and frequency-stabilized Nd:YAP/LBO (Nd-dropped YAlO$_3$%
/LiB$_3$O$_5$) laser (Yuguang Co. Ltd., FG-VIB) with both the harmonic-wave
output at $540$ $nm$ and the subharmonic-wave output at $1080$ $nm$ serves
as the laser source of the entanglement enhancement system. The output green
and infrared lasers are separated by a beam splitter M10 coated with high
reflection for $540$ $nm$ and high transmission for $1080$ $nm$. The green
laser serves as the pump fields of three NOPAs and the infrared laser is
used for the injected signal of NOPA, as well as the local oscillators (LO)
of BHD1 and BHD2. The traveling-wave mode cleaners (MC1 for $540$ $nm$ and
MC2 for $1080$ $nm$) consisting of three mirrors are used for the optical
low-pass filters of noises and the spatial mode cleaners. The finesses of
MC1 for $540$ $nm$ and MC2 for $1080$ $nm$ both are $550$. The optical
isolators (ISO1 for $540$ $nm$ and ISO2 for $1080$ $nm$) are utilized to
prevent the feedback optical fields from the NOPA returning to the laser. We
choose the $\alpha $-cut Type II KTP (KTiOPO$_4$) to be the nonlinear
mediums in the three NOPAs, which can achieve type-II noncritical
phase-matched frequency-down-conversion of the pump field at 1080 nm. The
size of the three KTP crystals is the same ($3*3*10mm^3$) and temperature
of the KTP crystal in the three NOPAs is controlled around $63^{\circ }C$ to
satisfy the phase-matching condition. Since the phase-matching temperature
of KTP has a broad full width of about $30^{\circ }C$, we can make the signal
and the idler modes double-resonating inside a NOPA by carefully tuning the
temperature of the crystal around $63^{\circ }C$. The NOPA1 is in a
semi-monolithic Fabry-Perot configuration consisting of a KTP crystal and a
concave mirror (M1) with $50$ $mm$ radius of curvature. The front face of
the crystal is coated to be used as the input coupler of the pump field (the
transmissivity of $99.8\%$ at $540$ $nm$ and $0.04\%$ at $1080$ $nm$), and
the other face is coated with the dual-band antireflection at both $540$ $nm$
and $1080$ $nm$. M1 coated with transmissivity of $5.2\%$ at $1080$ $nm$ and
high reflection at $540$ $nm$ is used as the output coupler and is mounted
on a piezoelectric transducer (PZT1) to scan actively the cavity length of
NOPA1 or lock it on resonance with the injected signal as needed. The length
and the finesse of the cavity of NOPA1 are $54$ $mm$ and $115$,
respectively. The NOPA2 (3) has the ring configuration consisting of two
flat mirrors M2 (6) and M3 (7) and two concave mirrors M4 (8) and M5 (9)
with $100$ $mm$ radius of curvature. The KTP crystal with the $1080$ $nm$
and $540$ $nm$ dual-band antireflection coated at both end faces is placed
in the middle of M4 (8) and M5 (9). M2 (6) serves as the input-output
coupler with the transmission of $10.0\%$ at $1080$ $nm$ and antireflection
at $540$ $nm$, respectively. All the other mirrors are high reflection at $%
1080$ $nm$ and antireflection at $540$ $nm$. M5 (9) are mounted on PZT2 (3)
for scanning or locking actively the length of the optical cavity NOPA2 (3). The
length and the finesse of the cavity for both NOPA2 and NOPA3 are $557$ $mm$
and $60$, respectively. The threshold pump powers of the three NOPAs are $250$
$mW$ for NOPA1 and $900$ $mW$ for NOPA2 and NOPA3, respectively. The signal
and the idler optical beams with orthogonal polarizations produced by NOPA3
are separated by a polarizing-beam-splitter (PBS) and then
are detected by BHD1 and BHD2, respectively. The BHD consists of a $50/50$ \
beam-splitter and a pair of photodiodes (ETX-500 InGaAs). Locking the
relative phase between the signal (idler) beam and the LO to 0 or $\pi /2$,
the fluctuations of amplitude or phase quadrature of the signal (idler) field
can be measured by BHD1 (BHD2). The noise powers of the amplitude (phase)
quadratures simultaneously measured by BHD1 and BHD2 are combined by the
positive (negative) power combiner ($\oplus (\ominus )$) and then the
correlation variances of the amplitude-sum (phase-difference) are analyzed
by a spectrum analyzer (SA).

\begin{figure}[tbp]
\begin{center}
\includegraphics[width=8.6cm]{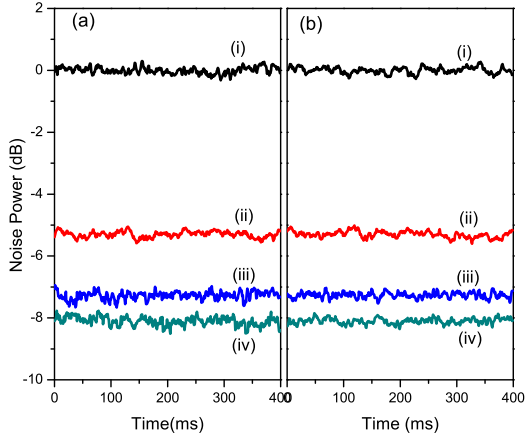}
\end{center}
\caption{(Color online) Noise powers of the correlation variances: (a) Trace
(i), the QNL; trace (ii), correlation variance of $\langle \delta ^2(\hat{X}%
_{a_{_1}}^{out}+\hat{X}_{a_{_2}}^{out})\rangle $; trace (iii), correlation
variance of $\langle \delta ^2(\hat{X}_{b_{_1}}^{out}+\hat{X}%
_{b_{_2}}^{out})\rangle $; trace (iv), correlation variance of $\langle \delta
^2(\hat{X}_{c_{_1}}^{out}+\hat{X}_{c_{_2}}^{out})\rangle $. (b) Trace (i), The
QNL; trace (ii), correlation variance of $\langle \delta ^2(\hat{Y}%
_{a_{_1}}^{out}-\hat{Y}_{a_{_2}}^{out})\rangle $; trace (iii), correlation
variance of $\langle \delta ^2(\hat{Y}_{b_{_1}}^{out}-\hat{Y}%
_{b_{_2}}^{out})\rangle $; trace (iv), correlation variance of $\langle \delta
^2(\hat{Y}_{c_{_1}}^{out}-\hat{Y}_{c_{_2}}^{out})\rangle $.}
\end{figure}

Figs. 4 (a) and (b) show the measured correlation variances of the
amplitude-sum and the phase-difference, respectively. In Fig. 4 (a) [(b)]
trace (i) is the QNL, trace (ii), (iii) and (iv) are the measured noise
power spectra of the amplitude-sum (phase-difference) at 2 MHz for EPR1,
EPR2 and EPR3, respectively. The initial correlation variances of EPR1
produced by NOPA1 are $\langle \delta ^2(\hat{X}_{a_{_1}}^{out}+\hat{X}%
_{a_{_2}}^{out})\rangle =$ $\langle \delta ^2(\hat{Y}_{a_{_1}}^{out}-\hat{Y}%
_{a_{_2}}^{out})\rangle =0.59$, corresponding to $-5.3\pm 0.2$ $dB$ below
the QNL. After the first-stage enhancement by NOPA2 the correlation variances are reduced to $\langle
\delta ^2(\hat{X}_{b_{_1}}^{out}+\hat{X}_{b_{_2}}^{out})\rangle =\langle
\delta ^2(\hat{Y}_{b_{_1}}^{out}-\hat{Y}_{b_{_2}}^{out})\rangle =0.38$,
corresponding to $-7.2\pm 0.2$ $dB$ below the QNL. At last, after the
cascaded enhancement by NOPA2 and NOPA3, the correlation variances of EPR3
become $\langle \delta ^2(\hat{X}_{c_{_1}}^{out}+\hat{X}_{c_{_2}}^{out})%
\rangle =\langle \delta ^2(\hat{Y}_{c_{_1}}^{out}-\hat{Y}_{c_{_2}}^{out})%
\rangle =0.31$, corresponding to $-8.1\pm 0.2$ $dB$ below the QNL. The
correlation variance of EPR3 is denoted in Fig. 2 (a) with a red star, where
$\gamma _1=0.1$, $\gamma _2=0.004$, $r=0.83$, ($-7.2$ $dB$ below the QNL)
corresponding to the operation conditions of NOPA3. On the trace (i) of Fig.
2 (b) we can see, when the correlation variance of the input EPR beam (EPR2)
equals to $-7.2$ $dB$, the correlation variance of the output field (EPR3)
is $-8.3$ $dB$ which is in good agreement with the experimental measured
value ($-8.1$ $dB$). We also measured the anti-correlation variances of EPR1, EPR2 and EPR3, which are $\langle \delta ^2(\hat{X}_{a_{_1}}^{out}-\hat{X}%
_{a_{_2}}^{out})\rangle =$ $\langle \delta ^2(\hat{Y}_{a_{_1}}^{out}+\hat{Y}%
_{a_{_2}}^{out})\rangle =13.6$, $\langle
\delta ^2(\hat{X}_{b_{_1}}^{out}-\hat{X}_{b_{_2}}^{out})\rangle =\langle
\delta ^2(\hat{Y}_{b_{_1}}^{out}+\hat{Y}_{b_{_2}}^{out})\rangle =25.9$, and $\langle \delta ^2(\hat{X}_{c_{_1}}^{out}-\hat{X}_{c_{_2}}^{out})%
\rangle =\langle \delta ^2(\hat{Y}_{c_{_1}}^{out}+\hat{Y}_{c_{_2}}^{out})%
\rangle =34.2$ corresponding to $8.3$ $dB$, $11.1$ $dB$ and $12.3$ $dB$ above
the QNL, respectively. The sums of the amplitude and the phase correlation variances for EPR1, EPR2 and EPR3 are $1.18$, $0.76$, and $0.62$, respectively, all of which satisfy the inseparability criterion, i.e. these values are smaller than $4$ (when the sum is larger than 4 the signal and the idler optical modes in the output field are separable and thus do not form an entangled state \cite{Duan,Simon}. )

In conclusion, we have experimentally demonstrated that the CV entanglement
of optical field can be enhanced by the cascaded NOPA. The upper limitation
of the enhanced entanglement depends on the intra-cavity loss ($%
\gamma _2$) of the NOPA and the relative phase fluctuation $\theta $ of the phase locking system. In our system $\gamma _2=0.4\%$ for NOPA3, if $%
\gamma _2$ can be reduced to $0.1\%$ as that reached by Ref. \cite{Mehmet},
the entanglement of EPR3 will increase to $-9.6$ $dB$ below the QNL. The
presented scheme opens an avenue to enhance CV entanglement using easily
reachable optical devices.

This research was supported by National Basic Research Program of China
(Grant No. 2010CB923103), Natural Science Foundation of China (Grants Nos.
60736040, 11074157 and 60821004), the TYAL.

\end{document}